# LOW-COMPLEXITY LSTM-ASSISTED BIT-FLIPPING ALGORITHM FOR SUCCESSIVE CANCELLATION LIST POLAR DECODER

*Chun-Hsiang Chen, Chieh-Fang Teng, and An-Yeu (Andy) Wu*

Graduate Institute of Electronics Engineering, National Taiwan University, Taipei, 106, Taiwan.
{johnny, jeff}@access.ee.ntu.edu.tw, andywu@ntu.edu.tw

## ABSTRACT

Polar codes have attracted much attention in the past decade due to their capacity-achieving performance. The higher decoding capacity is required for 5G and beyond 5G (B5G). Although the cyclic redundancy check (CRC)-assisted successive cancellation list bit-flipping (CA-SCLF) decoders have been developed to obtain a better performance, the solution to error bit correction (bit-flipping) problem is still imperfect and hard to design. In this work, we leverage the expert knowledge in communication systems and adopt deep learning (DL) technique to obtain the better solution. A low-complexity long short-term memory network (LSTM)-assisted CA-SCLF decoder is proposed to further improve the performance of conventional CA-SCLF and avoid complexity and memory overhead. Our test results show that we can effectively improve the BLER performance by 0.11dB compared to prior work and reduce the complexity and memory overhead by over 30% of the network.

## 1. INTRODUCTION

Polar codes, since proposed by Arikan in 2009, have attracted lots of attention from both academia and industry due to their capacity-achieving performance [1]. Further, in 2016, polar codes have been adopted as the official channel coding scheme for enhanced mobile broadband (eMBB) uplink and downlink control channels [2].

Recently, as deep learning (DL) has many revolutionary breakthroughs in the field of computer vision and natural language processing, many researchers have also been dedicated to applying this powerful technique to enhance decoding algorithms [3]-[7]. However, their decoding capacity is still worse than the state-of-the-art CRC-assisted successive cancellation list (CA-SCL) [8]-[10]. Although CA-SCL benefits from the list structure and CRC, it still suffers from large storage overhead and computational complexity. As the result, bit-flipping decoding was adopted in CA-SCLF to release hardware complexity with the flexible tradeoff between the decoding performance and latency as shown in Fig. 1(a) [11]-[13].

This research work is financially supported by the MediaTek Inc., Hsin-chu, Taiwan, under Grants MTKC-2019-0070.

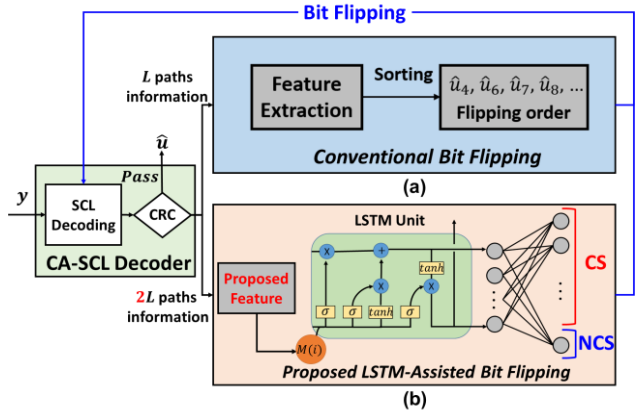

**Fig. 1.** Overview of (a) the conventional CA-SCLF design [13], and (b) the proposed low-complexity LSTM-assisted CA-SCLF.

To further improve decoding capacity and reduce the decoding latency, the flipping trials of error bits in CA-SCLF, we take advantage of DL to enhance the error path prediction accuracy. Besides, to avoid the severely increased memory overhead and additional computations from DL model, we apply our domain knowledge in communication systems with carefully designed DL architecture. The proposed long short-term memory network (LSTM)-assisted CA-SCLF is shown in Fig. 1(b). The main contributions of our work are as follows:

1) We apply LSTM to the CA-SCLF decoder, which can effectively extract the features from the sequential decoding process and enhance the error bit prediction. Therefore, the proposed design can improve the block error rate (BLER) performance and decoding latency by 0.11dB and 41.25%, respectively.

2) By taking advantage of domain knowledge for input data preprocessing and output dimension reduction, we can significantly reduce the memory requirements and computational complexity by over 30%, making our design more feasible for hardware implementation.

The rest of this paper is organized as follows. Section **2** briefly reviews the concept of polar codes, CA-SCLF, and the prior work. Section **3** illustrates the proposed architecture of LSTM-assisted CA-SCLF with novel input data preprocessing and output design. The numerical experiments and analyses are shown in Section **4**. Finally, Section **5** concludes our work.

## 2. PRELIMINARIES

### 2.1. Polar Codes

In this paper, we use $u_j^i$, with $j < i$, to denote the vector $\{u_j, u_{j+1}, \ldots, u_i\}$. To construct an $(N, K)$ polar codes, the mathematical foundations lay on the polarization effect of the $2 \times 2$ transformation matrix $\boldsymbol{F} = \begin{bmatrix} 1 & 0 \\ 1 & 1 \end{bmatrix}$. $N$ virtual bit-channels are constructed from recursive polarizing transformation by $\log_2 N$ times. As $N$ approaches infinity, the channels separate into the noisy channels (unreliable) and noiseless channels (reliable). According to the reliabilities of each channel, the $K$ information bits are assigned to the $K$ most reliable channels in the $N$-bit message $\boldsymbol{u}_1^N$ and the remaining $(N - K)$ bits are taken as frozen bits with the assignment of zeros. The set of information bit indices is denoted by $A$. Then, the $N$-bit transmitted codeword $\boldsymbol{x}_1^N$ is calculated by multiplying $\boldsymbol{u}_1^N$ with generator matrix $\boldsymbol{G}_N$ as follows:

$$\boldsymbol{x}_1^N = \boldsymbol{u}_1^N \boldsymbol{G}_N = \boldsymbol{u}_1^N \boldsymbol{F}^{\otimes n} \boldsymbol{B}_N, n = \log_2 N, \quad (1)$$

where $\boldsymbol{F}^{\otimes n}$ is the $n$-th Kronecker power of $\boldsymbol{F}$, and $\boldsymbol{B}_N$ represents the bit-reversal permutation matrix.

In past years, there are several decoding algorithms developed to meet various scenarios, such as belief propagation (BP), CRC-assisted successive cancellation bit flip (CA-SCF) [11], and CA-SCL. BP iteratively decodes the message over the corresponding factor graphs. On the other hand, both CA-SCF and CA-SCL decode the message based on SC but with the different mechanisms to correct the error decoding bits.

### 2.2. Prior Work: CA-SCLF Decoding [12]-[13]

CA-SCLF decoder is recently developed due to its high performance and flexibility. The decoding process of CA-SCLF is shown in Fig. 2. is described as the following. Firstly, given the received signal $\boldsymbol{y}$, the CA-SCL algorithm with determined list size $L$ is implemented. The concept of CA-SCL is to reserve the $L$ most reliable paths (red lines) from $2L$ temporary paths through the path metric (PM) values $PM_l^i$, where $l$ and $i$ denote the index of $2L$ temporary paths and the bit index. The path metric is defined as

$$PM_l^i = \sum_{j=1}^{i} \ln(1 + e^{-(1-2\hat{u}_j[l]) \cdot L^{(j)}[l]}), \quad (2)$$

where $\hat{u}_j[l]$ and $L^{(j)}[l]$ denote the $j$-th decoding bit at $l$-th path and the decoding LLR of bit $\hat{u}_j[l]$ given $\boldsymbol{y}$ and decoding trajectory $\hat{u}_1^{j-1}[l]$, respectively. Without loss of generality, we assume that the $2L$ temporary paths satisfy $PM_1^i < PM_2^i < \cdots < PM_{2L}^i$. Besides, $l'$ and $l'+L$ stand for the reserved paths (first $L$ paths) and discarded paths (last $L$ paths), respectively.

Secondly, CA-SCLF attempts to correct the first error path selection, where the correct path (orange arrow line)

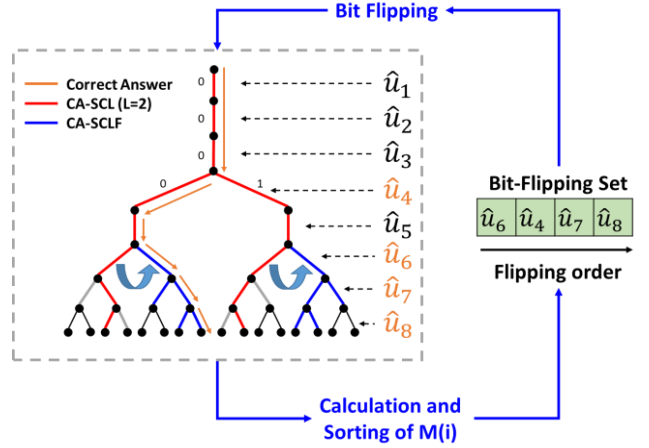

**Fig. 2.** Decoding procedure of CA-SCLF with $(N, K) = (8, 4)$.

is not among the reserved paths (red lines). In Fig. 2., the first error happens at $\hat{u}_6$. As the result, there will be a set for bit flipping (BFS). In [13], the BFS is based on the self-defined error metric $M(i)$:

$$M(i) = \sum_{l'=1}^{L} \prod_{j=1, j \in A}^{i} (1 + e^{-(1-2\hat{u}_j[l]) \cdot L^{(j)}[l]})^{-1} \quad (3)$$

Next, we retry CA-SCL from the discarded paths (blue lines) by following the bit-flipping order. We should note that there is only one bit flipped for every trial. The decoding continues until the CRC passes or the trial number has reached the maximum number $T_{max}$.

We can observe that the error metric from (3) only contains the information from discarded paths. It results in information loss from the reserved paths, which will degrade the performance. In our work, we look into the information from both the reserved paths and discards paths. Also, we are the first to utilize LSTM for CA-SCLF.

## 3. PROPOSED LOW COMPLEXITY LSTM ASSISTED CA-SCLF POLAR DEOCDER

### 3.1. Proposed Feature Extraction for Neural Network

From the original implementation of CA-SCL in the first step of CA-SCLF decoding, we can receive the LLRs and PM values for each decoding bit. However, if we set the whole LLRs or PM values as the neural network inputs, the cost of memory and complexity will be unaffordable. As we know, the path metric represents the probability of each decoding path given $\boldsymbol{y}$. As the result, we propose the novel error metric based on PM and be able to represent the error path selection probability. The formula takes the ratio of the discarded paths and reserved paths without any information loss and is written as

$$M(i) = \frac{\sum_{l'=1}^{L} e^{-(PM_{l'+L}^i)}}{\sum_{l'=1}^{L} e^{-(PM_{l'}^i)}}. \quad (4)$$

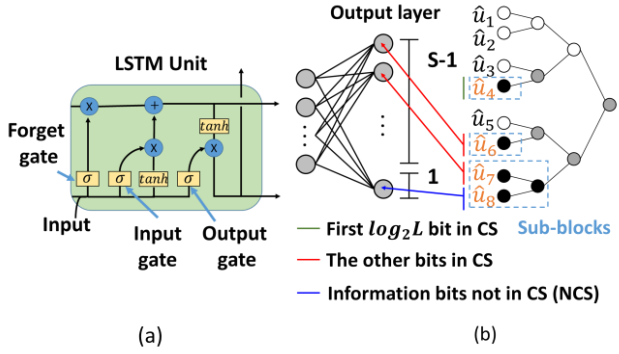

**Fig. 3.** Our proposed LSTM network is composed of the (a) LSTM unit and (b) fully connected layer with $(N, K) = (8, 4)$ and $L = 2$.

### 3.2. Proposed Low Complexity Feature Extraction for Neural Network

The exponential and division operation in Eq. (4) leads to high complexity which will limit the benefit. As the result, we modify the formula in the following. Firstly, we replace exponential function by the addition, and we will get

$$M(i) = \frac{\sum_{l'=1}^{L} e^{-(PM_{l'+L}^{i})}}{\sum_{l'=1}^{L} e^{-(PM_{l'}^{i})}} \approx \frac{L - (PM_{1+L}^{i} + PM_{2+L}^{i} + \cdots + PM_{l'+L}^{i})}{L - (PM_{1}^{i} + PM_{2}^{i} + \cdots + PM_{l'}^{i})}. \quad (5)$$

Eq. (5) derives from the Tylor series approximation. Next, we can take the logarithm operation from Eq. (5) since the logarithm operation causes no information loss and then approximate the logarithm function with the linear function. Therefore, we have

$$\tilde{M}(i) = \ln(1 - \sum_{l'=1}^{L} \frac{PM_{l'+L}^{i}}{L}) - \ln(1 - \sum_{l'=1}^{L} \frac{PM_{l'}^{i}}{L}) \quad (6)$$

$$\approx \sum_{l'=1}^{L} \frac{PM_{l'}^{i}}{L} - \sum_{l'=1}^{L} \frac{PM_{l'+L}^{i}}{L}. \quad (7)$$

Finally, we multiply Eq.(7) by $-L$ for simplicity, and there is no information loss. Hence, we will get

$$\hat{M}(i) = \sum_{l'=1}^{L} PM_{l'+L}^{i} - \sum_{l'=1}^{L} PM_{l'}^{i}. \quad (8)$$

From Eq. (8), we can see that the equation is rewritten in a hardware-friendly form.

### 3.3. Proposed Low Complexity LSTM Model for CA-SCLF Design

According to the sequential decoding process and long code length, the LSTM model is used in this work. The LSTM model is composed of a cell, an input gate, an output gate, and a forget gate as shown in Fig. 3(a).

Firstly, we calculate the modified error metrics of the last $K - \log_2(L)$ information bits from Eq. (8) to be the serial network inputs instead of the path metrics. As for the outputs, the fully connected network with softmax as its

**Algorithm 1:** Proposed LSTM-Assisted CA-SCLF Decoding

- **Input:** $y, L, T_{max}, A$
- **Output:** $\hat{u}_1^N$
- **Computation**

    **for** $i = 1, 2, \ldots, N$ **do**

        **Do the conventional CA-SCL**

        $\{\hat{u}_1^N, PM_l^i, PM_{l+L}^i\} \leftarrow CA - SCL(\emptyset)$

        **Calculate** $\hat{M}(i)$ by Eq. (8)

    **end**

**if** $CRC(\hat{u}_1^N) = $ success **then**

    **Return** $\hat{u}_1^N$

**else**

    Build the bit-flipping set (BFS) ← LSTM with input $\hat{M}(i)$

**end**

**for** $T = 1, 2, \ldots, T_{max}$ **do**

    Bit-flipping index $u_i \leftarrow BFS(T)$

    **Redo the CA-SCL from the discarded paths at** $u_i$

    $\{\hat{u}_1^N, \sim, \sim\} \leftarrow CA - SCL(i)$

    **if** $CRC(\hat{u}_1^N) = $ success **then**

        **Return** $\hat{u}_1^N$ and **break**

    **end**

**end**

activation function is connected with the LSTM. We take the concept of critical set (CS) for the outputs in our work. The CS consists of the bits which can be taken as the first bit in every sub-block of polar codes with code rate equal to 1. These bits are proven to have higher error rate [11]. As the result, we propose to use one-hot vector with the length $S$ whose first $S - 1$ bits represent the information bit indices in CS, excluding the first $\log_2(L)$ information bit indices, and the last bit represents the other information bit indices not in critical set (NCS) as shown in Fig. 3(b). Relying on the specific design of inputs and outputs, the complexity and memory can be extremely reduced. The loss function is cross-entropy between neural network output $\hat{o}$ and true label $o$, which is written as

$$L(\hat{o}, o) = \frac{-1}{S} \sum_{j=1}^{S} \hat{o}_j \log(o_j) + (1 - \hat{o}_j) \log(1 - o_j). \quad (9)$$

The decoding procedure is shown in Algorithm 1.

## 4. SIMULATION RESULTS

The simulation setup is summarized in Table I.

TABLE I. SIMULATION PARAMETERS

| | |
|---|---|
| Encoding | Polar code (512,256), CRC = 24, $S = 62$ |
| EbN0 (dB) | 0, 0.5, 1, 1.5, 2 |
| Training/Validation data size | 500,000/100,000 |
| Mini-batch size | 200 |
| Optimizer | Adam |
| Simulation Environment | DL library of MATLAB with NVIDIA RTX 2080 GPU |

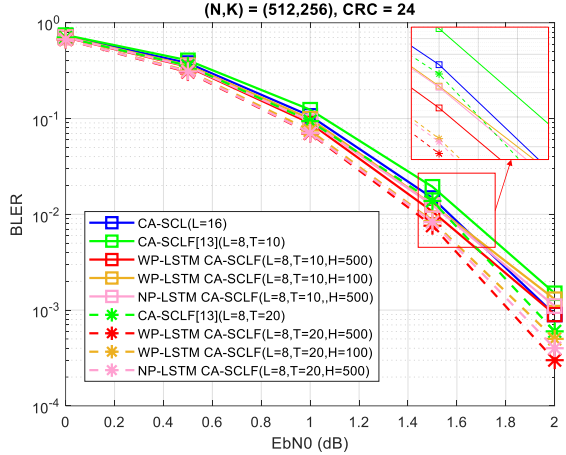

**Fig. 4.** Comparison of BLER performance between CA-SCL, CA-SCLF [13], and the proposed LSTM design.

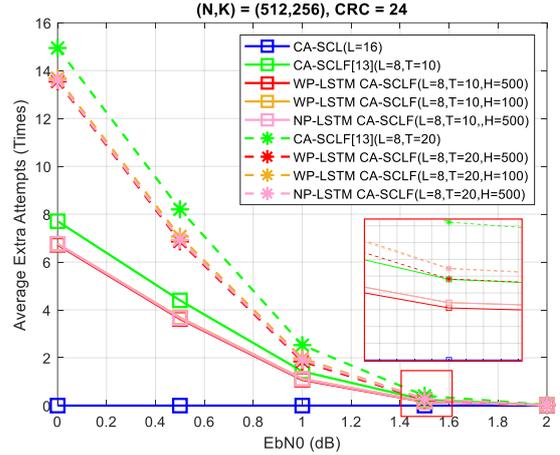

**Fig. 5.** Comparison of average extra attempts (latency) between CA-SCL, CA-SCLF [13], and the proposed LSTM design.

### 4.1. Performance of Proposed LSTM Design for CA-SCLF

In first experiment, we examine the block error rate (BLER) performance of the proposed LSTM design with preprocessing (WP) named as WP-LSTM. Similar to the output design of WP-LSTM with the path metrics taken as inputs, we also simulate for the performance of LSTM without preprocessing (NP) named as NP-LSTM. The proposed design is compared with the prior work [13], and the simulation results are shown in Fig. 4. Except for CA-SCL, all the others are implemented with $L = 8$. $T$ and $H$ denote the $T_{max}$ and hidden layer size in the figure, respectively.

Firstly, we can observe that WP-LSTM with $T_{max} = 10$ can outperform the CA-SCL decoder with $L = 16$ by 0.06dB and 0.13dB with $T_{max} = 20$. Also, WP-LSTM has 0.11dB better performance than [13] when $T_{max} = 20$. Secondly, we compare WP-LSTM with NP-LSTM with the same hidden layer size 500 and $T_{max} = 20$. The results show that WP-LSTM can achieve 0.018dB better performance, which tells the truth that the domain knowledge cannot only reduce the complexity and memory overhead but also help to improve the performance under limited resources. Furthermore, we reduce the hidden layer size of WP-LSTM to 100, we can observe that the performance just slightly degrades 0.024dB which means the design is robust.

In second experiment, we examine the average bit-flipping attempts (latency). The simulation results are shown in Fig. 5. We can observe that the average bit-flipping attempts of the proposed design are relatively small. Compared with the conventional design [13], the proposed design can reduce 41.25% of average extra attempts with $T_{max} = 20$ at EbN0 = 1.5dB.

From the simulations above, we can know that the proposed design cannot only enhance the BLER performance but also reduce the latency.

### 4.2. Complexity Analysis of Neural Network Design

In Table II, we compare the hardware complexity of the proposed design and the LSTM design without domain knowledge, whose inputs are the LLRs of $N$-bit message in $L$ paths and the outputs are the whole information set. The design rule of the latter design is similar to [7].

In Table II, the list size $L$ and hidden layer size are set to 8 and 100, respectively. We can see that by taking advantage of domain knowledge for input data preprocessing and output dimension reduction, the proposed LSTM design can effectively reduce about 32% of complexity and memory usage.

TABLE II. COMPLEXITY ANALYSIS OF DIFFERENT NETWORK DESIGN

|  | LSTM network w/o domain knowledge | Proposed LSTM network w/ domain knowledge |
|---|---|---|
| Addition | $\sim H(4H+4L+K+5)$<br>~69300 (100%) | $\sim H(4H+S+9)$<br>~47100 (67.9%) |
| Multiplication | $\sim H(4H+4L+K+3)$<br>~69100 (100%) | $\sim H(4H+S+7)$<br>~46900 (67.8%) |
| Memory | $\sim H(4H+4L+K+6)$<br>~69400 (100%) | $\sim H(4H+S+10)$<br>~47200 (68.0%) |

### 5. CONCLUSION

In this paper, we present the novel LSTM-assisted CA-SCLF decoder. The bit-flipping order can be learned from the LSTM network, and it can improve the performance and the latency compared to the conventional algorithms [13]. Also, we utilize the domain knowledge to reduce the complexity and memory requirements so that the hardware can be more friendly. In the future, we will research into the design for supporting multiple code rate.